# Kerker-Type Positional Disorder Immune Metasurfaces

Hao Song[1], Binbin Hong[2], Neng Wang[2] and Guo Ping Wang[2]*

[1]College of Physics and Electronic Information Engineering, Neijiang Normal University, Neijiang, 641100, P. R. China

[2] College of Electronics and Information Engineering, Shenzhen University, Shenzhen, 518060, P. R. China

*Email: gpwang@szu.edu.cn

**Abstract**

Metasurfaces that can work without the rigorous periodic arrangement of meta-atoms are highly desired by practical optical micro-nano devices. In this work, we proposed two kinds of Kerker-type metasurfaces possessing positional disorder immunity. The metasurfaces are composed of two different core-shell cylinders satisfying the first and second Kerker conditions, respectively. Even with large positional disorder perturbation of the meta-atoms, the metasurfaces can still maintain the same excellent performances as periodic ones, such as the total transmission and magnetic mirror responses. This disorder immunity is due to the unidirectional forward and backward scatterings of a single core-shell cylinder leading to very weak lateral couplings between neighboring cylinders thus rarely affecting the multiple scatterings in the forward or backward direction. In contrast, the dominant response of the disordered non-Kerker-type metasurface decreases significantly. Our findings provide a new idea for designing robust metasurfaces and extend the scope of metasurface applications in sensing and communication under complex practical circumstances.

**Keywords**: Kerker scattering, disorder immunity, metasurfaces, multiple scatterings, lateral coupling

## 1 Introduction

Two-dimensional (2D) planar metasurface with periodic sub-wavelength meta-atoms has been growing interest in the efficient optical integrated components applied in communication [1], national defense [2], biomedicine [3], and so on. Unlike the classical optical path accumulation in continuous media, metasurfaces can independently control each meta-atom's scattering amplitude, phase, and polarization through the Pancharatnan-Berry phase, localized surface plasmon resonance, and Mie resonance [4], resulting in the minimal propagation path of the light-matter interaction. Therefore, metasurfaces can tailor the electromagnetic (EM) wavefront at will in various frequencies due to introducing abrupt changes in EM responses, and have achieved the novel generalized Snell law [5,6], photon spin Hall effect [7], planar microlens [8], vector beam [9], holography [10], etc.

Most importantly, to obtain the predetermined EM responses, the meta-atoms that are regarded as the secondary sources should be arranged on the surface of a substrate in some specific global order such as periodic in general [5, 6, 11-16], although the azimuth, size, shape, and material of the meta-atom can be changed randomly in each unit cell [17, 18]. Obtaining perfect periodic subwavelength patterning structures requires more sophisticated and precise fabrication technologies, e. g. photolithography, electron-beam lithography, and focused-ion-beam lithography, which are time-consuming and expensive [4, 14, 19]. However, in practice, the meta-atoms will inevitably be disturbed disorderly by limitations of the precision of fabrication and some inevitable external forces in unstable and harsh environments, which will destroy the order arrangement and lead to failure to obtain the desired EM performances of metasurfaces [20-23]. Moreover, the periodicity leads to pronounced spatial dispersion and reduces the degree of freedom of manipulation [24, 25]. Therefore, reducing the dependence on periodicity is preferable for metasurfaces to exhibit the desired high performances in dealing with positional disordered perturbation, namely obtaining the disorder immune metasurfaces.

Then, if the positional disorder perturbation is introduced, the EM responses of metasurfaces would be changed compared with the periodic ones [26-28]. Thus, a key to disorder immune metasurfaces is how to minimize the unpredictable couplings (or interactions) caused by the disordered spacings between meta-atoms [29-31]. To solve this problem, a good candidate is the meta-atoms with exciting Mie resonances, which can confine the EM field around or inside the particles leading to the weak coupling between particles as long as they are not very close to [32, 33]. Additionally, as a special case, the non-radiative anapole mode can confine the electric field inside the particles [34-36], which can be applied to the disorder immune metasurfaces because of the vanished coupling between particles [37]. However, the resonant modes cannot deal with the case of small spacing, and the non-radiative anapole mode has no EM responses in the far field.

Encouragingly, the particles with unidirectional scattering can maintain the feature even in a random ensemble [38]. Therefore, in this paper, we proposed two Kerker-type disordered immune metasurfaces exhibiting total transmission and magnetic mirror responses, respectively. Kerker scattering was found by Kerker et al. when studying the scattering of the magnetic sphere [39]. Then, the first (second) kind of Kerker scattering is zero-backward (zero-forward) scattering originating from the destructive interferences of the equal contributions of the electric (ED) and magnetic (MD) dipoles in the backward (forward) direction resulting in the forward (backward) directional scattering [39-42]. Up to now, the related works have extended to the generalized Kerker effect [43-45], lattice Kerker effect [46, 47], acoustic Kerker effect [48], biological Kerker effect [49, 50], optomechanical Kerker effect [51], and so on. Moreover, the magnetic mirror is a new type of high reflector with an in-phase between reflected and incident electric field at the interface, which can enhance light-matter interactions, applied in subwavelength imaging [52], molecular fluorescence [53], Raman spectroscopy [54], perfect reflectors [55], etc. Furthermore, the Kerker-type metasurfaces can be employed for antenna engineering [11, 56] and reconfigurable intelligent surfaces [1, 57] to realize the low-loss and stable unidirectional signal transmitting and receiving and achieve high-precision positioning.

Here, we design two meta-atoms: the hollow $Al_2O_3$ cylinder supports the first kind of Kerker scattering, while the copper-$Al_2O_3$ manifests the second kind of Kerker scattering. The very weak lateral scatterings rarely affect the multiple scatterings of the meta-atoms. Despite the positional disordered perturbation, the random Kerker-type metasurfaces maintain the total transmission and magnetic mirror responses unchanged. In contrast, the performance of the random non-Kerker-type metasurface has decreased significantly due to the unpredictable lateral scattering couplings. Furthermore, our finding is also suited for unidirectional scattering in other angles beyond the forward and backward. Therefore, the Kerker-type metasurfaces can be used as high-robustness metasurfaces, show the potential of the single-step fabricated metasurfaces, and enhance the adaptability of metasurfaces in various practice environments for sensing, wireless communications, and photonics applications.

## 2 Results and discussions
### 2.1 Disorder immunity of the first kind of Kerker metasurface

We begin by providing a meta-atom composed of an infinite-long hollow cylinder in Fig. 1(a). The inner and outer radii of the cylinder are $r_0$ and $r_1$, respectively. The incident plane electromagnetic (EM) wave is polarized with the electric field **E** parallel to the axis of the cylinder and the incident frequency is 10 GHz. The shell is made of $Al_2O_3$ with relative permittivity $\varepsilon_1 = 9.424 + i2.920 \times 10^{-3}$ [58] at the given frequency. Here, we fix the size parameter $x_r = kr_1$ as 1.343, where $k$ is the wavenumber in the air. For simplicity, we introduce the dimensionless parameter $\alpha = r_0/r_1$.

Then, we analyze the scattering properties of an individual cylinder. The semi-analytical scattering multipole decomposition using Cartesian bases is based on displacement current density, which is expressed as [59]

$$\mathbf{J}(\mathbf{r}) = -i\omega\varepsilon_0[\varepsilon(\mathbf{r}) - 1]\mathbf{E}(\mathbf{r}), \tag{1}$$

where $\omega$ is the incident angular frequency, **r** is the position vector, $\varepsilon_0$ is vacuum permittivity, $\varepsilon(\mathbf{r})$ is relative permittivity, and $\mathbf{E}(\mathbf{r})$ is the total electric field. Thus, the basic Cartesian moments [60-63] of the ED, MD, and magnetic quadrupole (MQ) are expressed as

$$p_j = \frac{1}{-i\omega}\int \mathbf{J}(\mathbf{r})\mathrm{d}^2 r, \tag{2}$$

$$m_j = \frac{1}{2c}\int \mathbf{r} \times \mathbf{J}(\mathbf{r})\mathrm{d}^2 r, \tag{3}$$

$$Q_{jp}^m = \frac{1}{3c}\int \{[\mathbf{r} \times \mathbf{J}(\mathbf{r})]_j r_p + [\mathbf{r} \times \mathbf{J}(\mathbf{r})]_p r_j\}\mathrm{d}^2 r, \tag{4}$$

where $c$ is the speed of light in vacuum, subscripts $j$ and $p$ are components of **r**, $r=|\mathbf{r}|$. The contributions of higher-order EM modes are vanishing small and safely ignored [63, 64]. On the other hand, the toroidal moments of the ED, MD, and MQ are given by

$$T_j^e = \frac{1}{10c}\int \{[\mathbf{r} \cdot \mathbf{J}(\mathbf{r})]r_j - 2r^2 J_j\}\mathrm{d}^2 r, \tag{5}$$

$$T_j^m = \frac{ik}{20c}\int [\mathbf{r} \times \mathbf{J}(\mathbf{r})]_j r^2 \mathrm{d}^2 r, \tag{6}$$

$$T_{jp}^{mq} = \frac{ik}{42c}\int r^2 \{r_j[\mathbf{r} \times \mathbf{J}(\mathbf{r})]_p + [\mathbf{r} \times \mathbf{J}(\mathbf{r})]_j r_p\}\mathrm{d}^2 r. \tag{7}$$

Subsequently, a total multipole moment $(\hat{S} + ik\hat{T}^{(S)})$ is calculated by the interference between the base Cartesian moment $\hat{S}$ and its toroidal moment $\hat{T}^{(S)}$. The scattering contribution of a multipole is proportional to $|\hat{S} + ik\hat{T}^{(S)}|^2$ [61].

Based on Eqs. 1-7, the scattering efficiencies and the summation of all modes are shown in Fig. 1(b). At point A ($\alpha=0.185$), the scattering efficiencies of ED and MD are equal, and the MQ is very weak. According to the spectral region of point A, the induced ED and MD modes are in phase, which will suppress the radiation in backward direction [41, 42].

With the induced multipole moments of the ED, MD, and MQ, the angular scattering intensity reads as [61, 65, 66]

$$\mathbf{I}_{SA}(\theta) = f_0 \left| -\mathbf{P} - \mathbf{M}\cos\theta + \frac{ik}{2}\mathbf{Q}^m \cos 2\theta \right|^2 \tag{8}$$

where $\theta$ is the scattering angle and the forward and backward directions correspond to 0 and 180 degrees, respectively. $f_0$ is a constant when $k$ and incident **E** are fixed. The **P**, **M**, and **Q**$^m$ are the total multipole moments of the ED, MD, and MQ modes, respectively. The angular scattering pattern of each mode and their superposition (sca) are demonstrated in Fig. 1(c). The interference between the ED and MD leads to significant directional scattering in the forward direction where the lateral and backward scatterings are suppressed. The weak backward scattering is due to the presence of the MQ mode [67]. Nevertheless, the forward-to-backward ratio of the sca is still as high as 4.63. The asymmetry parameter [68]

$$g = \frac{\oiint_s \cos\theta W_{sca}(\theta) ds}{\oiint_s W_{sca}(\theta) ds} \tag{9}$$

is calculated to demonstrate the degree of the unidirectional, where $W_{sca}$ is the scattering energy, $\cos\theta = \mathbf{n}\cdot\hat{\mathbf{k}}$ and **n** is the vector normal to the surface $s$ enclosing the particle. $g=0$ is the light scattering isotropically or laterally, $g=1$ is the unidirectional forward scattering, and $g=-1$ is the unidirectional backward scattering. Here $g=0.83$ indicates the dominant scattering in the forward direction. Moreover, assume the parameter $\beta$ is the ratio of the dominant scattering peak to the lateral scattering of 90 degrees. Here, $\beta=17.62$, which indicates the lateral scattering is very weak. As a result, the cylinder exhibits the first kind of Kerker scattering approximately [43, 45].

We design a metasurface composed of the core-shell cylinders arranged periodically along the *x*-direction with the minimal face spacing $p$ depicted in Fig. 1(d). The metasurface is normally illuminated by a plane EM wave with a *z*-polarized **E**. In this paper, the commercial software COMSOL [69] is used. The periodic boundary condition is applied along the *x*-direction and the perfectly matched layer (PML) is used to replace the outermost air along the *y*-direction. When $p$ changes from $0.05\lambda$ to $0.45\lambda$ with an interval of $0.05\lambda$, Fig. 1(e) depicts the transmissivity (T) spectra of the metasurfaces as functions of $\alpha$ and $p$. Interestingly, peaks of all the transmissivity spectra have identical values (about 0.99) near point A. Therefore, at point A where the first kind of Kerker condition is approximately fulfilled, the metasurface sustains total transmission for a wide range of $p$.

We truncate the periodic metasurface to only 51 cylinders with $p=0.25\lambda$, and the reflectivity (Rp) and transmissivity (Tp) of the finite-size periodic metasurface at point A are shown in Fig. 1(f). For the simulation of the finite-size structures, PMLs are introduced in all directions. The metasurface still exhibits the nearly total transmission due to the Tp being about 0.99 and Rp=0. Then, the positions of cylinders are perturbed with $p$ varying from $0.05\lambda$ to $0.45\lambda$ randomly. The disorder degree ($\sigma$) of the metasurface is defined as $\sigma = \max\left(\left|\Delta p_i / p_0\right|\right)$, where $\Delta p_i = p_i - p_0$ with $p_i$ being the $p$ of arbitrary adjacent cylinders and $p_0$ is the initial $p$. We prepare 10 different random metasurfaces with $\sigma=80\%$. The Rr and Tr are the reflectivity and transmissivity of the random samples, respectively. The average value of all Tr is mT about 0.94 decreasing only 0.05 compared to the Tp, while the average value of all Rr is mR still zero. Meanwhile, the red dots are located on the red line with a deviation of about 0.0055. Therefore, the transmission of random metasurfaces has strong stability.

Figure 2(a) shows the far-field radiations of the finite-size periodic (per) and random (ran) metasurfaces. The dominant EM responses are both the zeroth-order transmission, and the two peaks are almost identical because the radiations in other angles are very weak. Figures 2(b) and 2(c) depict the **E** distributions of the periodic and random metasurfaces, respectively. The transmitted wavefront is perturbed weakly due to the appearance of weak radiations in the non-zero-degree directions, notwithstanding the random metasurface still maintains almost total transmission.

Next, we discuss the multiple scatterings of the cylinders at point A. Figure 3(a) shows the angular scattering patterns of two cylinders when $p$ varies from $0.05\lambda$ to $0.45\lambda$. The lateral scatterings of all curves are very weak. Despite the weak coupling slightly decreasing the magnitude of peaks when the particles are very close, the curve shape of the unidirectional forward scattering of the two cylinders is robust to the variation of $p$. Figure 3(b) depicts the multiple scatterings of N cylinders with $p=0.25\lambda$. All patterns show the unidirectional forward scattering and suppression of the backward and lateral scatterings. Moreover, the magnitude of the peak is approximately proportional to N satisfying the grating diffraction [70]. Therefore, the disorder immunity of the metasurfaces originates from the weak lateral scattering of the cylinders with exciting the first kind of Kerker scattering, which leads to the negligible lateral scattering coupling that rarely affects the multiple scatterings.

**2.2 Disorder immunity of the second kind of Kerker metasurface**

Now, we switch to the second kind of Kerker metasurface. The meta-atom is an infinite-long copper-$Al_2O_3$ core-shell cylinder in Fig. 4(a), and the perfect electric conductor (PEC) can displace the copper at the incident frequency. Under the plane EM wave, Fig. 4(b) shows the scattering efficiencies of total scattering (sca), ED, and MD modes of the isolated cylinder, where $x_r$ is 0.785. The other weak higher-order modes are neglected. Point B denotes $\alpha=0.432$ and the equal contributions of ED and MD. According to the spectral region of point B, the induced ED and MD modes are anti-phase leading to scattering cancellation in the forward direction [42].

Figure 4(c) displays the angular scattering of the individual cylinder at point B. The interference of ED and MD induces strong total scattering (sca) in the backward direction, whereas the forward and lateral scatterings nearly vanish. The incident wave and the nonzero imaginary part of $\varepsilon_1$ hinder complete suppression of the forward scattering, while forward scattering can disappear in the gain material [40]. Nonetheless, the backward-to-forward scattering ratio is as high as 11.25, and $g=-0.60$. Here, $\beta=4.87$, which indicates the lateral scattering is still weak. Therefore, the cylinder excites the unidirectional backward scattering satisfying the second kind of Kerker scattering nearly [45].

The schematic of a total reflective periodic metasurface composed of cylinders is shown in Fig. 4(d). The metal substrate can ensure total reflection. Figure 4(e) demonstrates the phase difference ($\Delta\phi_r$) between the reflected and incident electric field at the top tangent plane of the metasurface as $p$ varying from $p_0$ ($0.05\lambda$) to $p_1$ ($0.45\lambda$) in a step of $0.05\lambda$. All curves change continuously from negative to positive values and intersect at point B where $\Delta\phi_r=0$. Therefore, the metasurface at point B is a magnetic mirror within a wide change range of $p$.

Figure 4(f) shows the EM responses of finite-size periodic and random metasurfaces at point B. The reflectivity (Rp) and $\Delta\phi_r$ (Pp) of the periodic metasurface with $p=0.25\lambda$ are 0.986 and -0.00911 rad, respectively. Hence, the finite-size periodic metasurface still exhibits the magnetic mirror response. Then, we prepare 10 random metasurfaces when $\sigma=80\%$ and $p_i$ is randomly in the range from $0.05\lambda$ to $0.45\lambda$. The Pr is the $\Delta\phi_r$ of the random metasurface and the mP is the average of all Pr. Here, mR=0.984 and mP=-0.00887 rad. The Rp, Rr, and mR have almost identical high reflectivity, and blue data maintain the $\Delta\phi_r$ about 0. As a result, the random metasurfaces still exhibit the magnetic mirror response.

We compare the far-field radiations of the finite-size periodic (per) and random (ran) metasurfaces at point B in Fig. 5(a). The two curves overlap perfectly indicating the random and periodic metasurfaces excite the same zeroth-order total reflection. Figure 5(b) shows the |**E**| distribution of the periodic

metasurface at point B, and Fig. 5(c) is the case of a random one. All metasurfaces exhibit the magnetic mirror response with a large-scale strongest interference electric field at their top interfaces. The perturbation of the wavefronts comes from the diffuse reflection at the side boundaries, while the random case also includes the interference of weak non-zeroth order diffractions. Therefore, the magnetic mirror response of the second kind of Kerker metasurface is robust to large disordered position perturbation.

We control the $\Delta\phi_r$ of the phase curves' intersection point of the infinitely periodic metasurface as a function of $x_r$ varying $p$ from 0.05λ to 0.45λ in Fig. 5(d). The underlying physical mechanisms of all points are the same as point B. However, the point will disappear due to the exciting mode change when $x_r$ is out of the range in this figure. The line with a slope of -1.267 is the fitting of the points. $\Delta\phi_r$ of the intersection point decreases linearly as $x_r$ increases. Point B is just located on the line. Figure 5(e) depicts the metasurface can be regarded as a PEC substrate and a dielectric film with a height of $2r_1$ effectively. Under the normally incident plane wave, the $\Delta\phi_r$ at the top interface of the structure reads as $\Delta\phi_r = 4kn_{eff}r_1$, where $n_{eff}$ is the effective refractive index. According to the slope, the $n_{eff}$ is 0.995. Therefore, the cylinders in the metasurface are equivalent to an air-like film for any non-invisible passive systems, and the $\Delta\phi_r$ is linearly adjustable by the thickness. Thus, we can linearly adjust the reflected phase of the metasurface by manipulating $x_r$, despite $p$ changes within a wide range. The controllable linear reflected phase is also suited for positional disorder immune gradient metasurfaces to achieve the reflection deflection, carpet cloak, etc [5, 37].

The multiple scatterings of two cylinders at point B as the $p$ varies from 0.05λ to 0.45λ are shown in Fig. 6(a). All patterns display strong backward scattering and suppressed forward and lateral scatterings. The main lobe beamwidth and magnitude decrease at a large $p$, which indicates the weak coupling vanishes. Figure 6(b) depicts the multiple scatterings of N cylinders at point B with $p$=0.25λ. Consequently, the cylinders still exhibit high unidirectional backward scattering. Therefore, the disorder immunity of the metasurface arises from the weak lateral scattering of the cylinder with exciting the second kind of Kerker scattering, which leads to very weak lateral scattering coupling.

## 2.3 Disorder immunity of the non-Kerker-type metasurface

In this section, we focus on the non-Kerker-type metasurface case. Figure 7(a) shows the meta-atom is an $Al_2O_3$ cylinder. Under the same incident wave, Fig. 7(b) demonstrates the scattering efficiencies of the total scattering (sca), ED, MD, and MQ modes. These modes have equal contributions at point C with $x_r$=1.15. In Fig. 7(c), the angular scattering of the sca shows the non-unidirectional scatterings of the cylinder at point C. Here, $g$=0.41 indicates more scattering power in the forward direction. $\beta$=3.66 lower than the Kerker metasurfaces values, indicating the lateral scattering becomes stronger.

Figure 7(d) demonstrates the schematic of the periodic metasurfaces. The reflectivities of the metasurface as the functions of $x_r$ and $p$ are shown in Fig. 7(e), where $p$ varies from 0.25λ to 0.51λ with a step of 0.01λ and the arrow denotes the increasing direction. Interestingly, all curves nearly overlap at point C with R>0.95. Therefore, the metasurfaces at point C have almost the same high reflectivities in that range of $p$.

Figure 7(f) depicts the reflectivity and transmissivity of the finite-size metasurfaces at point C. The periodic metasurface with $p$=0.38λ has Rp=0.976 and Tp=0. Then, 10 random metasurfaces with σ=34.21% are achieved by disordered distributing each $p$ of the periodic metasurface in the range of 0.25λ to 0.51λ. The distributions of Rr and Tr indicate the random metasurfaces are still high reflection without transmission. However, the mR=0.817 decreased by 0.159 compared with the Rp, and the deviation of all Rr is about 0.0141. Therefore, compared to the periodic metasurface, the response of the random metasurface falls off and fluctuates significantly.

The far-field radiations of the finite-size periodic and random metasurfaces are shown in Fig. 8(a). Both show dominant zeroth-order reflection but the radiations in other directions of the random metasurface are more evident. Hence, the decrease of the Rr is mainly ascribed to the increased radiations in other directions. The |**E**| distribution of the periodic metasurface in Fig. 8(b) indicates the high reflection originates from the waveguide-array (WGA) modes [71] and the lateral coupling between adjacent cylinders. Then, the |**E**| distribution of the random metasurfaces in Fig. 8(c) demonstrates the lateral couplings of the adjacent cylinders still occur but their strengths are different. Furthermore, the radiations of non-backward directions lead to the reflected wavefront perturbation and the speckles below the random metasurface.

We explore the multiple scatterings of two cylinders at point C in Fig. 9(a). The change of $p$ not only changes the magnitude and main lobe width of the forward and backward scatterings but also produces different lateral scatterings, then inducing different lateral scattering couplings between two cylinders. The various lateral couplings cause the non-backward radiations of the random metasurfaces and lead to the energy loss of the zeroth-order reflection and increased instability. Figure 9(b) demonstrates the multiple scatterings of N cylinders with $p=0.38\lambda$. The scatterings maintain the dominant forward and backward scatterings. Therefore, the non-Kerker metasurface has a lower disorder immune degree.

**Conclusion**

In summary, we have proposed two kinds of Kerker-type metasurfaces exhibiting total transmission and magnetic mirror responses immune to the large positional disorder perturbations of the meta-atoms. The meta-atoms of the two Kerker-type metasurfaces satisfy the first and second kinds of Kerker conditions, respectively. Because of the vanishing lateral scattering, the lateral couplings between meta-atoms are weak. As a result, the dominant EM responses of Kerker-type metasurfaces are highly robust to large positional disorder perturbations. In contrast, for the non-Kerker-type metasurface, the positional disorder will affect the non-ignorable lateral coupling between meta-atoms and thus decrease the dominant EM responses.

This mechanism can be applied to design other positional disorder immune metasurfaces. In addition, the high unidirectional scattering is not limited to the forward and backward directions of ED and MD modes and also can be extended to other directions by optimizing high-order electric and magnetic modes, e. g. achieving the generalized Kerker scattering [43]. This work paves a new way for designing highly robust metasurfaces and sheds new light on the practically accessible metasurfaces that suit complex circumstances. In future work, we will try to explore the Kerker-type metasurfaces immune to the incident polarization and angle by breaking the rotational symmetry of the meta-atoms [72].

**Acknowledgments**

This work is supported by the Key Project of the National Key R\&D program of China (No. 2022YFA1404500), the National Natural Science Foundation of China (No. 12074267), the Guangdong Basic and Applied Fundamental Research Foundation (No. 2020A1515111037), and the Shenzhen Fundamental Research Program (No. 20200814113625003).

**Disclosures**

The authors declare no competing financial interest.

**Figures**

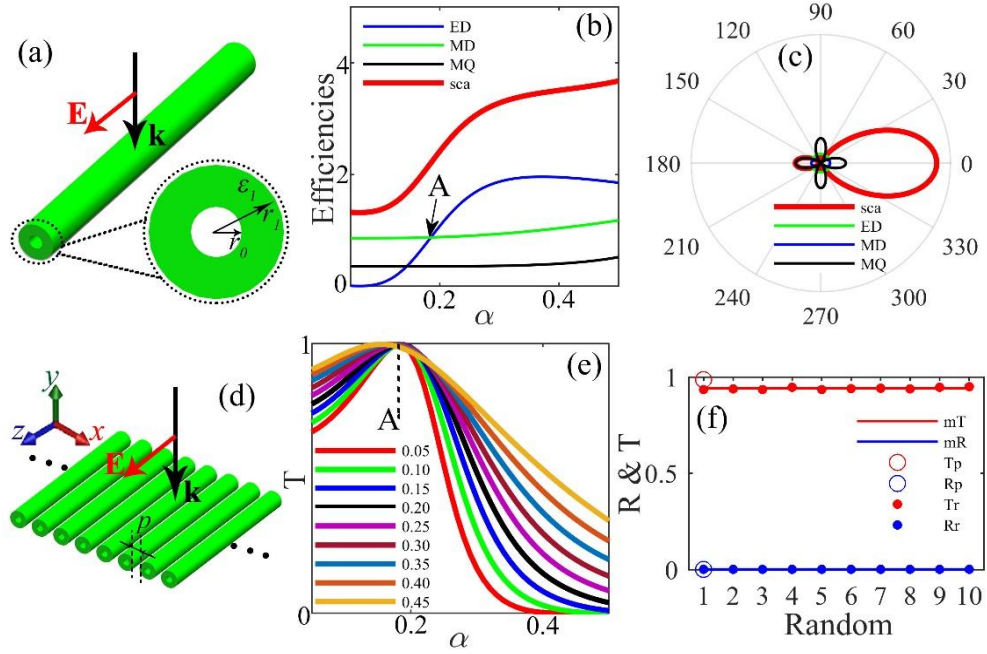

Figure 1. The first kind of Kerker-type metasurface. (a) Schematic of scattering analysis of an individual cylinder. (b) Scattering efficiencies of a single cylinder varying $\alpha$. The red curve (sca) is the total scattering. Point A denotes $\alpha=0.185$. (c) Angular scattering of point A. (d) Schematic of periodic metasurface with minimal face spacing $p$ under the normally incident plane wave with wavevector **k**. (e) Transmissivity (T) responses of the metasurfaces concerning $\alpha$. Each curve has a fixed $p$, which varies from $0.05\lambda$ to $0.45\lambda$ with an interval of $0.05\lambda$. (f) Reflectivity (R) and T responses of the finite-size metasurfaces at point A. Rp (Tp) is the R (T) of the periodic metasurfaces. Rr (Tr) is the R (T) of the random metasurfaces. The mR (mT) is the average value of all Rr (Tr).

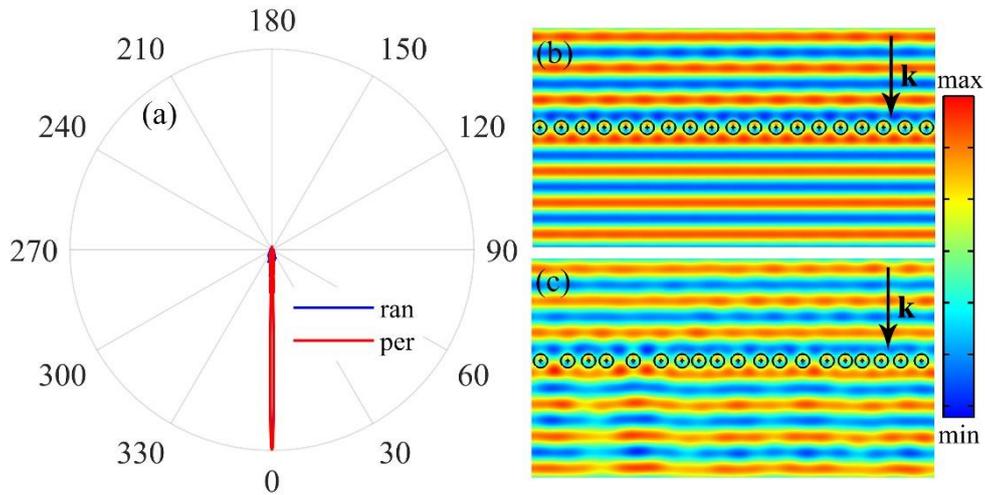

Figure 2. Responses comparisons of the periodic (per) and random (ran) metasurfaces at point A. (a) The far-field |**E**| of the per and ran metasurfaces. (b) and (c) The **E** distributions of the per and ran metasurfaces, respectively.

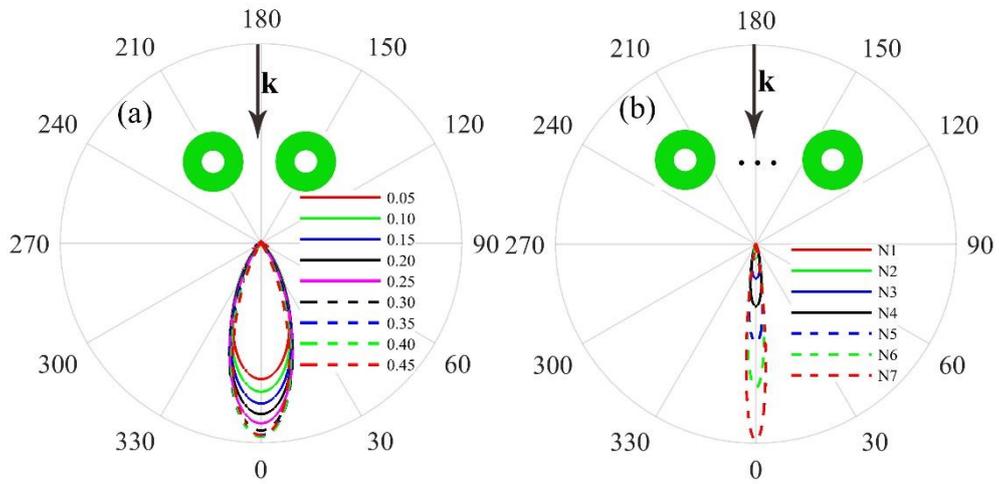

Figure 3. Multiple scatterings of the cylinders at point A. (a) Angular scattering patterns of the two cylinders varying $p$ from $0.05\lambda$ to $0.45\lambda$. (b) Angular scattering of N cylinders with fixed $p=0.25\lambda$. N denotes the number of cylinders increasing from 1 to 7.

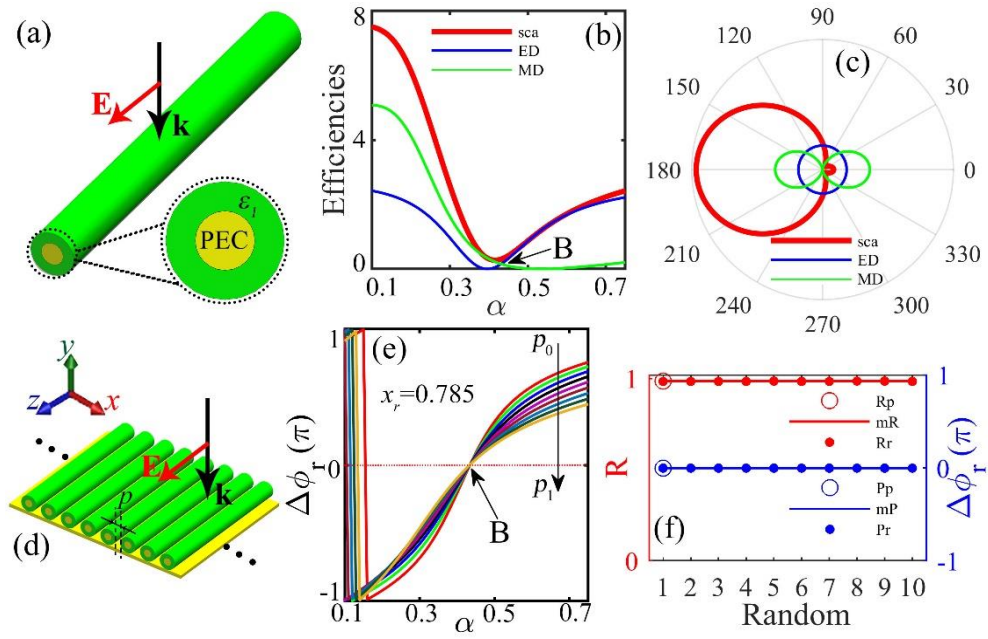

Figure 4. The second kind of Kerker-type metasurface. (a) Schematic of scattering analysis of an individual cylinder. (b) Scattering efficiencies of a single cylinder. Point B denotes $\alpha=0.432$. (c) Angular scattering of the cylinder at point B. (d) Schematic of the periodic metasurface. (e) The phase difference ($\Delta\phi_r$) of the reflected and incident electric field on the top interface plane of the metasurface when $x_r=0.785$ as $\alpha$ varies. Each curve has a fixed $p$ changing from $p_0$ (0.05$\lambda$) to $p_1$ (0.45$\lambda$) with a step of 0.05$\lambda$. (f) R and $\Delta\phi_r$ responses of the finite-size metasurfaces at point B. Pp: the $\Delta\phi_r$ of the periodic metasurface; Pr: the $\Delta\phi_r$ of the random metasurface; mP: the average value of all Pr.

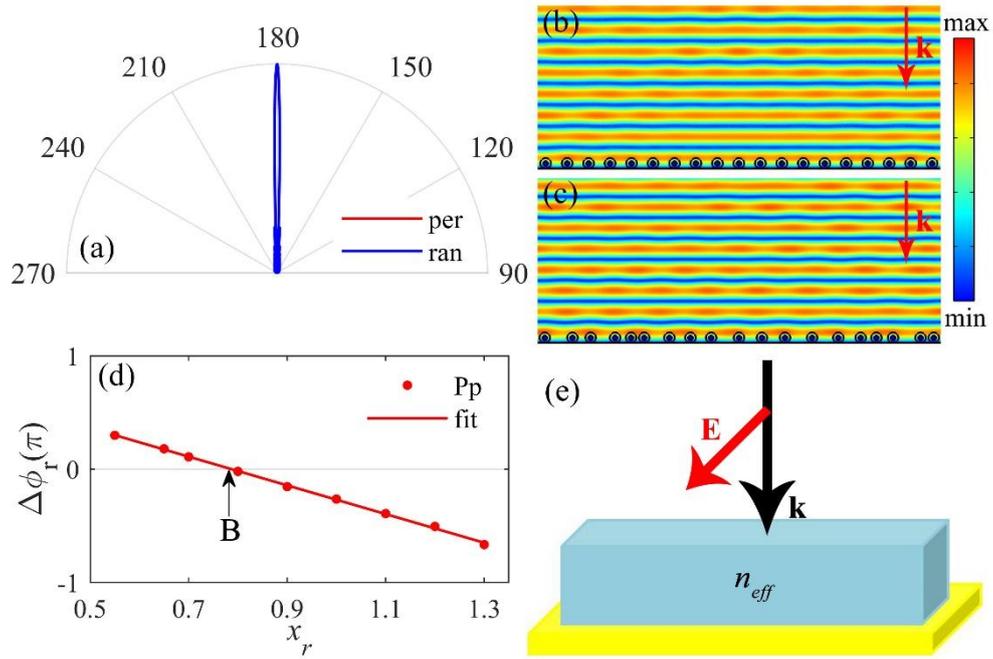

Figure 5. (a) The far-field |**E**| of the periodic and random metasurfaces at point B. (b) and (c) The fragments of |**E**| distributions of the periodic and random metasurfaces at point B, respectively. (d) Intersection point (see point B in Fig. 4(e)) of the $\Delta\phi_r$ as a function of $x_r$. The acquisition method is as same as that described in Fig. 4(e). The red line is the fitting of all data. (e) Schematic of the effective structure of the metasurface with the same incident wave, the thickness equals the external diameter of the cylinder.

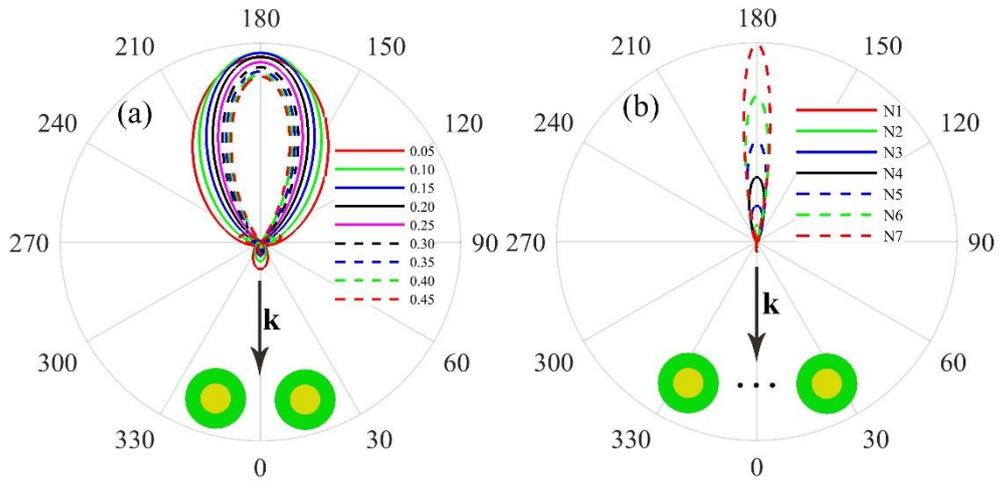

Figure 6. Multiple scatterings of the cylinders at point B. (a) Angular scatterings of two cylinders as *p* varying. (b) Angular scatterings of N cylinders with fixed *p*=0.25λ.

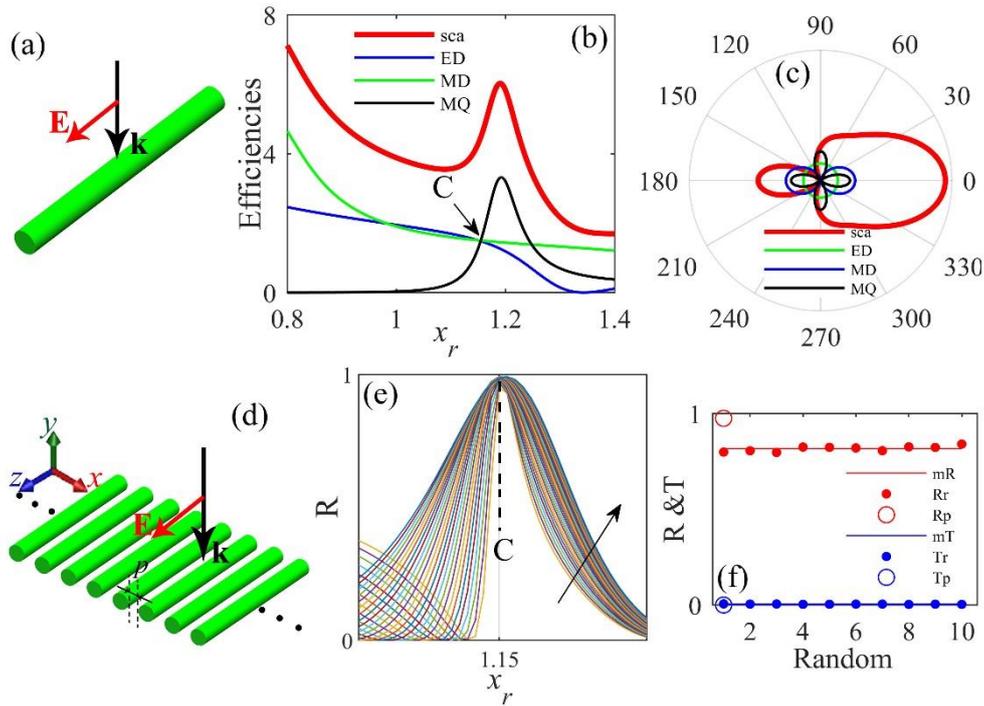

Figure 7. One kind of non-Kerker-type metasurface. (a) Schematic of scattering analysis of an individual cylinder with permittivity $\varepsilon_1$. (b) Scattering efficiencies of a single cylinder. Point C denotes $x_r$=1.15. (c) Angular scattering pattern of the cylinder at point C. (d) Schematic of the periodic metasurface. (e) The R responses of the metasurfaces concerning $x_r$. Each curve has a fixed *p* and the arrow denotes the *p* increasing from 0.25λ to 0.51λ. (f) R and T responses of the finite-size periodic and random metasurfaces at point C.

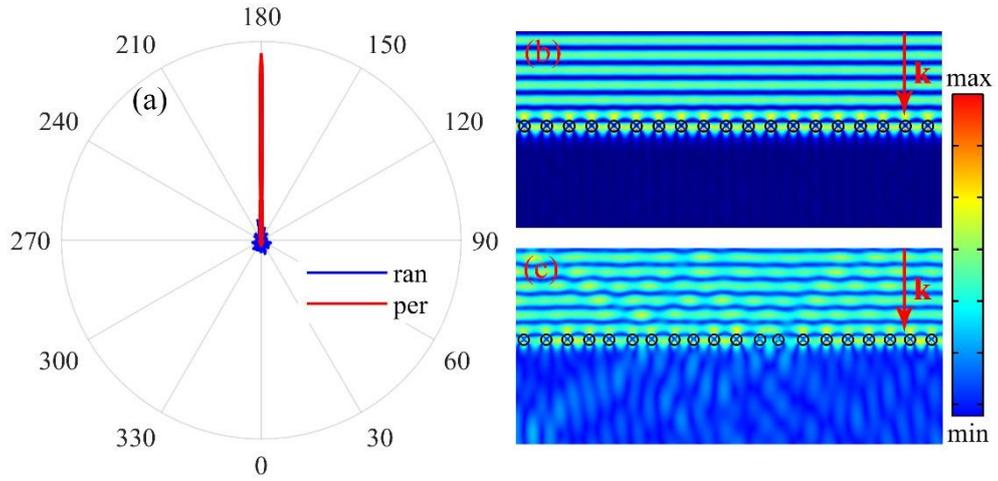

Figure 8. Response comparisons of the periodic and random metasurfaces at point C. (a) The far-field |**E**| of the per and ran metasurfaces. (b) and (c) electric field |**E**| distributions of the per and ran metasurfaces, respectively.

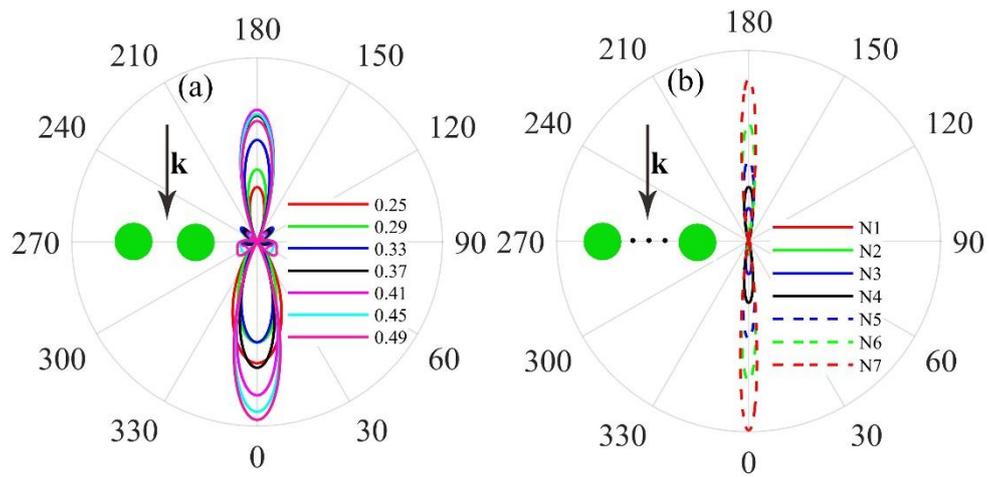

Figure 9. Multiple scatterings of the cylinders at point C. (a) Angular scattering patterns of the two cylinders varying $p$ from $0.25\lambda$ to $0.49\lambda$. (b) Angular scattering of N cylinders with fixed $p=0.38\lambda$.